\documentclass[conference]{IEEEtran}
\IEEEoverridecommandlockouts

\usepackage{cite}
\usepackage{amsmath,amssymb,amsfonts}
\usepackage{algorithmic}
\usepackage{graphicx}
\usepackage{physics}
\usepackage{textcomp}
\usepackage{booktabs}
\usepackage{xcolor}
\def\BibTeX{{\rm B\kern-.05em{\sc i\kern-.025em b}\kern-.08em
    T\kern-.1667em\lower.7ex\hbox{E}\kern-.125emX}}
\begin{document}

\title{On Quantum Random Walks in Biomolecular Networks\\
}

\author{%
  \IEEEauthorblockN{%
    Viacheslav Dubovitskii\IEEEauthorrefmark{1}\textsuperscript{\textsection},
    Aritra Bose\IEEEauthorrefmark{2}\textsuperscript{\textsection},
    Filippo Utro\IEEEauthorrefmark{2} and
    Laxmi Parida\IEEEauthorrefmark{2}
  }%
  \IEEEauthorblockA{\IEEEauthorrefmark{1} Algorithmiq Ltd, Helsinki, Finland}%
  \IEEEauthorblockA{\IEEEauthorrefmark{2} Quantum for Healthcare and Life Sciences, IBM Research, Yorktown Heights, NY}%
}
\maketitle
\begingroup\renewcommand\thefootnote{\textsection}
\footnotetext{Equal contribution}
\endgroup

\begin{abstract}
Biomolecular networks, such as protein–protein interactions, gene–gene associations, and cell–cell interactions, offer valuable insights into the complex organization of biological systems. These networks are key to understanding cellular functions, disease mechanisms, and identifying therapeutic targets. However, their analysis is challenged by the high dimensionality, heterogeneity, and sparsity of multi-omics data. Random walk algorithms are widely used to propagate information through disease modules, helping to identify disease-associated genes and uncover relevant biological pathways. In this work, we investigate the limitations of classical random walks and explore the potential of quantum random walks (QRWs) for biomolecular network analysis. We evaluate QRWs in two network-based applications. First, in a gene–gene interaction network associated with asthma, autism, and schizophrenia, QRWs more accurately rank disease-associated genes compared to classical methods. Second, in a structured multi-partite cell–cell interaction network derived from mouse brown adipose tissue, QRWs identify key driver genes in malignant cells that are overlooked by classical random walks. Our findings suggest that quantum random walks offer a promising alternative to classical approaches, with improved sensitivity to network structure and better performance in identifying biologically relevant features. This highlights their potential in advancing network medicine and systems biology.
\end{abstract}


\section{Introduction}
Biological networks provide a powerful framework for studying diseases by highlighting the interconnected nature of biological systems, rather than focusing on isolated molecular defects. At the core of this approach are network propagation algorithms~\cite{Cowen2017} which enable discovery of key disease mechanisms. Indeed, network propagation approaches operates on the assumption that genes or other biomolecules involved in similar biological function tend to interact within the same network. Therefore, genes are linked to a specific phenotype and their interactions can be mapped to a network. Subsequently, this information is iteratively spread across the network's edges to neighboring nodes, for a pre-determined number of steps or until a convergence is reached, effectively amplifying the signal associated to the phenotype. 

Random walks are one of the most commonly used methods for selecting biomolecules, studying drug-target interactions, and stratifying patients based on a disease-related interactome~\cite{Navlakha2010}. Recently, with the advent of quantum algorithms in pre-fault tolerant quantum hardware and their applications in biomedicine~\cite{burch2025towards}, clinical trials~\cite{doga2024can}, and single-cell analyses~\cite{basu2023towards}, among others. We aim to understand the limitations of classical random walks (CRW) and explore the possibilities of its quantum analogue, quantum random walks (QRW).  
Quantum random walks play an instrumental role in network medicine for link prediction, identifying disease modules in interactomes, and network propagation, among other tasks~\cite{maniscalco2022quantumnetworkmedicinerethinking, e25050730, Saarinen2024, utro2024perspective, Kempe_2003}. 
In particular, Saarinen et al.~\cite{Saarinen2024} propose a novel disease gene prioritization approach based on continuous-time quantum walks (CTQW) using the adjacency matrix of a protein–protein interaction (PPI) network. The CTQW algorithm was used to score candidate genes based on the idea that a protein is more likely to be associated with a disease if the walker is likely to transition from  that protein to any of the seed proteins. This is the only previous work that involves CTQW.

In this study, we demonstrate the practicality and versatility of quantum random walks on publicly available real-world biological networks with respect to its classical counterpart. In particular, we consider two main scenarios: (a) biomolecular Networks (b) cell-cell Interaction Networks (CCI). In the former, we are evaluating disease gene prediction ability of the two models on three different neurodegenerative disorders and several networks. While in the latter, we applied the two walkers to the CCI network in order to disentangle intricate interactions between cells, coordinate activities, and biological functions.



\section{Methods}
\label{sec:method}
\subsection{Biomolecular Network}
Consider a general biomolecular network represented by an undirected graph $G=(V,E)$, where $V$ is the set of biomolecules (genes, proteins, cells, etc.) represented as vertices in the graph and interactions between them captured by the set of edges $E$. The adjacency matrix, $A \in \mathbb{R}^{n\times n}$, for $n$ vertices of this graph is defined by

\begin{equation*}
    A = A_{ij} = \begin{cases}
        1 \textbf{ if $(i,j) \in E$}\\
        0 \textbf{ if $(i,j) \notin E$}
    \end{cases}
\end{equation*}

Then, the degree matrix $D\in\mathbb{R}^{n \times n}$ is defined as $D_{ii} = \sum_{j=1}^{n} A_{ij} \text{ } \forall \text{ } i \neq j$, and $D_{ij} = 0$. The Laplacian matrix $L \in \mathbb{R}^{n\times n}$ of graph $G$ is defined as $L = D-A$. 

\subsection{Cell-cell Interaction Network}
In a cell-cell interaction (CCI) network, the network can be represented as a multipartite directed graph $G_m = (V_m, E_m)$, where the set of vertices $V_m$ is partitioned into disjoint subsets 
$$
V_m = C_s \cup C_t \cup L \cup R
$$
where $C_s$ and $C_r$ are the set of vertices denoting sender and target receiver cells, and $L$ and $R$ are the set of ligands and receptors, respectively. We note that $C_s \cap C_r \cap L \cap R = \emptyset$, where $\emptyset$ is the null set. The edges represent interactions between biomolecules such as $(c_s, l) \in E$ when the sender cell type $c_s \in C_s$ interacts with the ligand $l \in L$. Similarly, $(l,r)$ captures the interaction between ligand $l$ and receptor $r \in R$. Lastly, the receptor $r$ and the receiver cell $c_t \in C_t$ interacts with the edge $(r,c_t)$. A valid communication path is represented as a three-hop path, $c_s \rightarrow l \rightarrow r \rightarrow c_t$. This naturally defines a four-partite graph, with no edges within any set of vertices, but between adjacent sets of vertices, as shown above. 

\subsection{Datasets}
\label{ssec:datasets}
\subsubsection{Biomolecular Networks}
To evaluate our approach, we consider different networks from publicly available repositories. In particular, we analyzed a network derived from summary statistics of genome-wide association studies (GWAS) on asthma, autism spectrum disorder (ASD), and schizophrenia in a variety of network propagation settings as indicated in~\cite{10.1093/bib/bbae014}. Five molecular networks were downloaded. There were HumanNet-XC, which combines protein-protein interactions (PPI) and gene functional information~\cite{kim2022humannet}; BioPlex3, containing two PPI networks derived from human embryonic kidney and colorectal carcinoma cell lines~\cite{huttlin2021dual}; ProteomeHD, a co-regulation map of human proteome~\cite{kustatscher2019co}; Parsimonious Composite network or PCNet derived from merging several biomolecular networks~\cite{huang2018systematic}; and STRING, which combines information from high-throughput sequencing, literature, and genomic context analysis~\cite{von2005string}. 

\subsubsection{Cell-Cell Interaction Networks}
To better understand cell–cell interactions (CCI) involving metabolites and sensor genes using DTQRW, we built a multi-partite CCI network from single-cell RNA-seq data of mouse brown adipose tissue (BAT)~\cite{Zheng2022.05.30.494067}. This allowed us to explore how myofibroblasts and malignant cells influence adipocytes through metabolic signaling and sensor genes such as solute carrier (SLC) transporters, which help move substances across cell membranes.

\subsection{Random Walk with Restart}
In the classical random walk with restart (RWR) framework, each walk on an undirected graph $G$, represented by a normalized adjacency matrix $A$, has a certain probability of returning to the root node. The walk propagates through the network, periodically restarting at the seed nodes. If the network is connected and the eigenvalues ($\lambda$) of $A$ follow $\lambda \leq 1$, this process converges to a steady-state that can be calculated as: 
\begin{equation}
    p_s = (1-\alpha)(I - \alpha A)^{-1}p_0 
\end{equation}

where $p_0$ in this case is a probability vector, defined by the measured gene-level (or proteins, SNPs, etc.) scores which are computed during network construction. The hyperparameter $\alpha$ is fixed, and the final probabilities of the steady state are used as scores to rank significance of the genes in a disease module. By ranking the top genes, the algorithm aims to prioritize disease-associated genes. The goal is to achieve high ranks for ground truth disease genes. 

\subsection{Continuous-time Quantum Random Walk}
A classical continuous-time random walk (CTRW) on a graph $G$ is defined by a walker that moves randomly between the vertices $V$ of the graph with random waiting times at each node governed by a continuous probability distribution. For a graph $G$ with $n$ vertices ($|V| = n)$, the classical walker uses a continuous-time Markovian process with a vector of probabilities $p$ to make the jump to a neighboring vertex, guided by 
$$
\frac{d}{dt}p = -L p
$$
for the Laplacian matrix $L$.  This evolution is directly contrasted with the Schr\"{o}dinger equation 
\begin{equation}\label{eq:schro}
\frac{d}{dt}\ket{\psi} = -iH\ket{\psi}
\end{equation}
where the quantum state is a vector of amplitudes $\psi \in \mathbb{C}^n$. This governs the evolution of the continuous-time quantum random walk (CTQRW) and causes the quantum walker to be distinctly different from the CTRW~\cite{aharonov1993quantum}. A CTQRW on a graph $G$ is defined by representing the adjacency matrix $A$ or Laplacian matrix $L$ as the Hamiltonian $H$ of the graph and setting up a transport problem such as the walker starts from a vertex $j$ and evaluates its projection on a target vertex $k$ following a Schr\"{o}dinger evolution (Equation~\eqref{eq:schro}) generated by the Hamiltonian $H$. Thus, in a CTQRW the probability $p$ of the walker moving from vertex $j$ to $k$, across time $t$ is 
\begin{align}\label{eq:proH}
    p(t) &= |\bra{j}e^{-iHt}\ket{k}|^2 \\
    \frac{d}{dt}p_{j->k}(t) &= 2\text{Re}\big(\bra{j}e^{-iHt}\ket{k}\bra{k}iHe^{iHt}\ket{j}\big)
    \end{align}
    
The unitary transformation $e^{-iHt}$ is based on the structure of the network captured by the Hamiltonian $H=L$ or $H=A$. In general, $H$ can be any Hermitian matrix related to $G$ that describes the structure of the network~\cite{annoni2024enhanced}. Considering this unitary transformation, the initial state vector with time $t$ can evolve as 
\begin{equation}
    \ket{\psi(t)} = e^{-iHt} \ket{\psi(0)}
\end{equation}

 CTQRW allows the walker to be in multiple states simultaneously due to quantum superposition. This property is used to prepare the initial state of the walker, defining the probabilities of finding the walker in several nodes proportionally to their scores, as if the walker started from a weighted probability distribution. Then, to prepare the initial state ($\psi(0)$) the scores have to be normalized before the evolution:
$$
\psi(0) = \frac{S}{\|S\|},
$$
where $S$ is a vector of gene scores. To prevent the walker to have a uniform probability distribution, the collapse of the wave function in quantum mechanics can be leveraged by periodically reinitializing the intermediate state based on the transition probabilities at the moment of collapse. 
$$
\psi(t) = \frac{\left|\psi(t_{\theta})\right|^2}{\|\left|\psi(t_{\theta})\right|^2\|}
$$
for any given state $t_\theta$.

To obtain a probability transition matrix from $H$, we evolve the system for a time $t$ and perform a measurement, which can by done using Equation~\eqref{eq:proH}, where $i = \sqrt{-1}$. Note that over long periods ($t\rightarrow \infty$), convergence to a steady state is not guaranteed in the same way as classical diffusion processes.

When $H$ is a real matrix it has been shown that transport from $j$ to $k$ cannot be directional with $\psi(t)$ as inverting the initial site and the target site leads to the same transport probability~\cite{annoni2024enhanced}. To overcome this, a general construction of \textit{chiral} quantum walk has been proposed~\cite{lu2016chiral}, which keeps the condition that the square modulus of the off-diagonal entries in $H$ is either 0 or 1, depending on connectivity. The transition probabilities are still defined by Equation~\eqref{eq:proH} but $H$ is no longer considered to be a real matrix and referred as chiral Hamiltonian of $G$ and $|H_{jk}| = A_{jk}$ and 
\begin{equation} 
H_{jk} = H^\dagger_{kj} = e^{i\phi_{jk}}
\end{equation}
when $\phi_{jk} \in [0,2\pi)$, so that in each edge $(j,k)$ in $E$, a complex phase $e^{i\phi_{jk}}$ is assigned.

When considering a specific disease in the disease gene prioritization task, we do not need the whole matrix exponential but rather its action on the initial state vector $\ket{\psi(0)}$. Consequently, calculating the scores can be an efficient process~\cite{FISCHER2017141}.

\subsection{Discrete-time Quantum Random Walk}
In a discrete-time classical random walk (DTRW) a walker jumps from one state to two possible states based on the outcome of a coin toss~\cite{xia2019random}. We can design an analogue of DTRW by associating a complex Hilbert space with the system defined by the graph $G$. The walker needs a node-dependent coin operator as the degree of each vertex (node) is different. Hence, the quantum state in a graph $G$ is defined as 
\begin{equation} \label{eq:dtqrw}
    \ket{\psi(t)} = \sum_{j=1}^{n}\sum_{k=1}^{d_j} \psi_{j,k} (t) \ket{j\rightarrow k}
\end{equation}
where $n$ is defined as the number of vertices, as above and $\ket{j\rightarrow k}$ is the state residing on vertex $j$ and is about to hop on to vertex $k$, and $d_j$ is the degree of the vertex $j$. The transition probability in this case is defined as 
\begin{equation} \label{eq:prodtq}
    p_{j->k}(t) = \sum_{k=1}^{d_j} |\psi_{j,k}(t)|^2
\end{equation}
 and the time evolution based on the Markov process is given by 
 \begin{align*}
      \ket{\psi(t)} &= U\ket{\psi(t-1)} \\
                    &= U^t \ket{\psi(0)}
 \end{align*}
where the unitary operator $U$ is the product of the shift operator $S$ and a coin operator $C$ and $U = SC$. The coin operator performs a unitary transformation on the coin state, creating a superposition of possible directions. Then, the shift operator $S$ is defined by the mapping 
$$
S\ket{j\rightarrow k} = \ket{k \rightarrow j}
$$
A key difference between classical and quantum discrete-time random walks is interference. In DTRW, the walker moves stochastically and in DTQRW it is deterministic and its path is reversible. The state is prepared by the superposition of amplitudes and has a faster spread, allowing for interference among the paths, where they combine constructively with higher probability, when multiple paths lead to the same node, or destructively, when amplitudes cancel and many nodes are reached with very low probabilities.

\subsection{Performance Metrics}
Evaluating disease gene prediction using network propagation methods deals with ranking genes in order of their importance found by random walk. Thus, a metric such as Average Precision at $K$ ($AP@K$), an ordered ranking metric that evaluates the relevance of the top $K$ predictions, as well as their positions in the ranking can be particularly useful. $AP@K$ is derived from precision at $K$ ($P@K$) measuring how many of the first $K$ items predicted are actually relevant with respect to a ground truth. Hence, 
$$
P@K = \frac{\text{Number of relevant genes in the first } K \text{ ranks}}{K}
$$
Therefore, $AP@K$ is defined as the average of $P@N$, where $N$ ranges from 1 to $K$. Formally, 
\begin{equation}
    AP@K = \frac{1}{min(K,M)} \sum_{N=1}^K P@N * rel(N)
\end{equation}
where $rel(N)$ is a binary indicator variable which assumes the value 1 if the $N^{th}$ ranked gene is a true positive and 0 otherwise.

\section{Results}
\label{sec:results}
\subsection{Biomolecular Networks}
Network propagation of biomolecular networks can be used to identify a list of candidate genes and translate them to score-based ranking, which can be an effective tool in biomarker discovery~\cite{Cowen2017}. Network propagation assumes interactions between biomolecules often captured by epistasis or linkage disequilibrium. It propagates prior information about genes associated with a phenotype or trait of interest through its nodes to their neighbors in an iterative manner until convergence or a predefined number of steps. Random walks are an effective way to investigate and represent network propagation~\cite{tong2008random}.
\begin{figure}[!htbp]
    \centering
    \includegraphics[width=\linewidth]{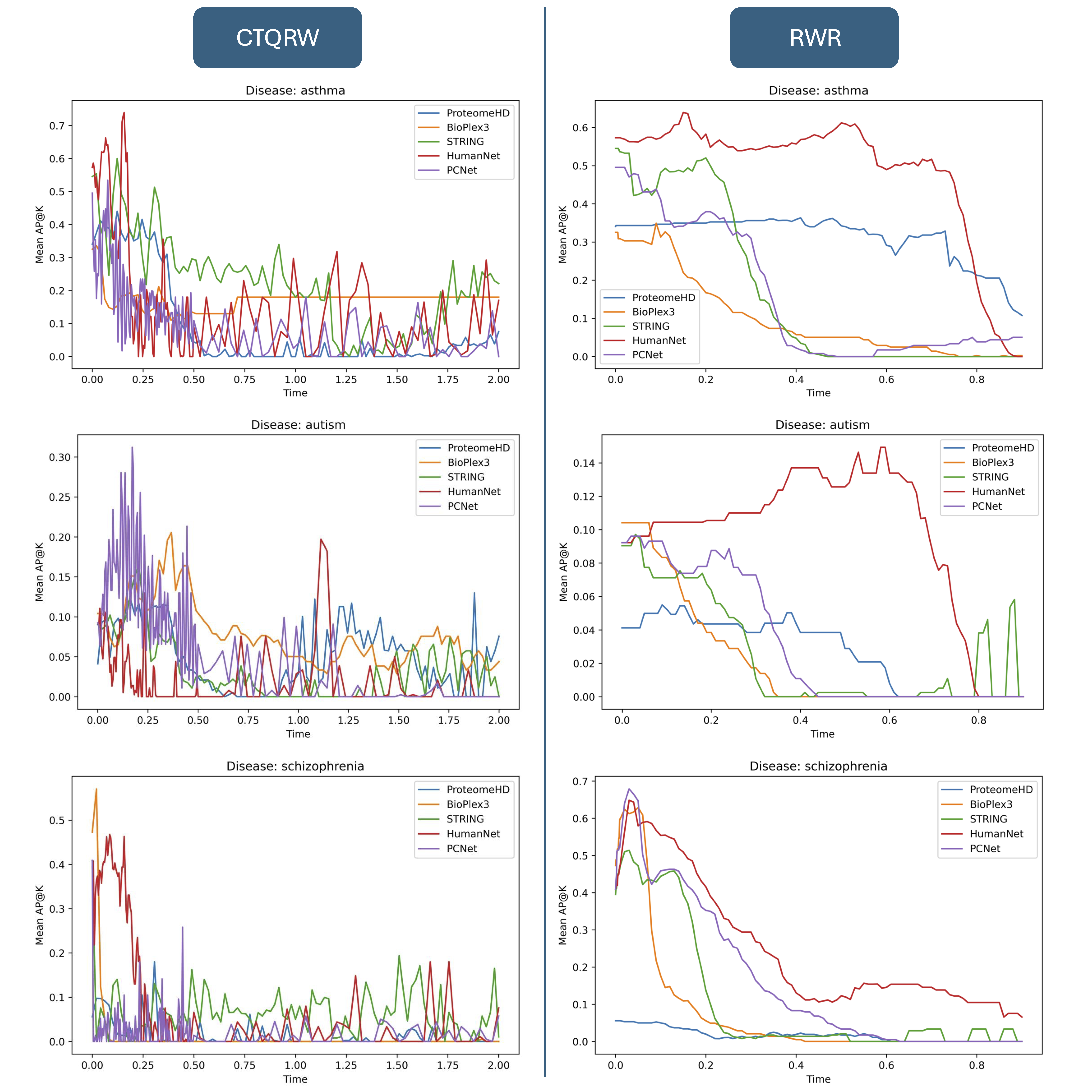}
    \caption{Mean $AP@20$ performance for Asthma, Autism, and Schizophrenia comparing CTQRW and RWR with varying time. }
    \label{fig:gwas_line}
\end{figure}
Network connectivity is essential for algorithms such as random walks or quantum random walks, as they rely on traversing the network to spread information. Notably, three networks: ProteomeHD, BioPlex3, and STRING resulted in fragments with many disconnected components or subgraphs (Table~\ref{tab:netstat}). These networks prevent propagation algorithms from effectively propagating information across isolated fragments. On the other hand, the HumanNet and the PCNet networks were connected and devoid of fragments. In our analysis, we extracted the greatest component (GC) from considered networks to apply the random walks.

We followed the steps to aggregate the GWAS targets from the GWAS catalog as shown in Vison\`{a} et al.~\cite{10.1093/bib/bbae014}. We define \textit{seed} genes that result statistically significant from the aggregated $p$-values of single nucleotide polymorphisms (SNP) to genes using PEGASUS~\cite{nakka2016gene}. A nominal significance level of $p < 0.01$ was used to select them. These \textit{seed} genes represent prior knowledge such that a link to the disease analyzed in the network, already exists. Target genes are considered from GWAS catalog with $p < 5\times 10^{-8}$, which is known to be an accepted threshold in literature for statistically significant SNPs associated with the disease. Next, we perform network propagation of the five downloaded networks namely, PCNet, BioPlex3, ProteomeHD, HumanNet, and STRING starting from the seed genes and propagating until we reach the target genes. We remove the genes at the intersection of seed and target genes, from the pool of targets. 
\begin{table}[!htbp]
\centering
   \resizebox{\columnwidth}{!}{%
    \begin{tabular}{l|l|l|l|l|l}
    \toprule
        \textbf{Network} & \textbf{Nodes} & \textbf{Edges} & \textbf{Fragments} & \textbf{GC Nodes} & \textbf{GC Edges} \\
        \midrule
        ProteomeHD & 2718 & 63,290 & 101 & 2471 & 62,598 \\
        Bioplex3 & 8364 & 35,704 & 99 & 8124 & 35,548 \\
        STRING & 17,079 & 420,534 & 95 & 16,855 & 418,069 \\
        HumanNet & 18,593 & 1,125,494 & 1 & 18,593 & 1,125,494 \\
        PCNet & 20,517 & 2,693,250 & 1 & 20,517 & 2,693,250 \\
        \bottomrule
    \end{tabular}
    }
    \caption{Number of nodes and edges of the biomolecular network and the greatest component (GC) of the network, respectively, along with the number of disconnected components or fragments in the networks.}
    \label{tab:netstat}
\end{table}

The connectivity of disease modules, which are subgraphs consisting of seed and target genes, differs across the five networks considered (Table~\ref{tab:gcstat}). This variation affects the accuracy of network propagation algorithms because some nodes forming a path from a seed node to a target node may lack biological significance and should ideally be excluded. We observe that the BioPlex3 network for all the diseases had the least relative size of GC with seeds and targets, whereas the STRING network had the largest among the networks with fragments.  

\begin{table}[!htbp]
\centering
\begin{tabular}{l|c|c|c}
\toprule
\textbf{Network} & \textbf{Asthma} & \textbf{ASD} & \textbf{Schizophrenia} \\
\midrule
ProteomeHD & 0.36 & 0.14 & 0.38 \\
BioPlex3   & 0.09 & 0.09 & 0.07 \\
STRING     & 0.61 & 0.36 & 0.62 \\
HumanNet   & 0.95 & 0.93 & 0.98 \\
PCNet      & 0.91 & 0.90 & 0.94 \\
\bottomrule
\end{tabular}
\caption{Relative size of giant component of disease module (seeds + targets)}\label{tab:gcstat}
\end{table}

By evaluating $AP@K$ for each network propagation experiment in these networks, we recovered target genes. Then we applied CTQRW on these networks and compared its performance with the PageRank implementation of random walk with restart (RWR) which is a common classical random walk technique from the python package \texttt{NetworkX}~\cite{hagberg2008exploring}. 

\begin{figure*}[h]
    \centering
    \includegraphics[width=0.75\linewidth]{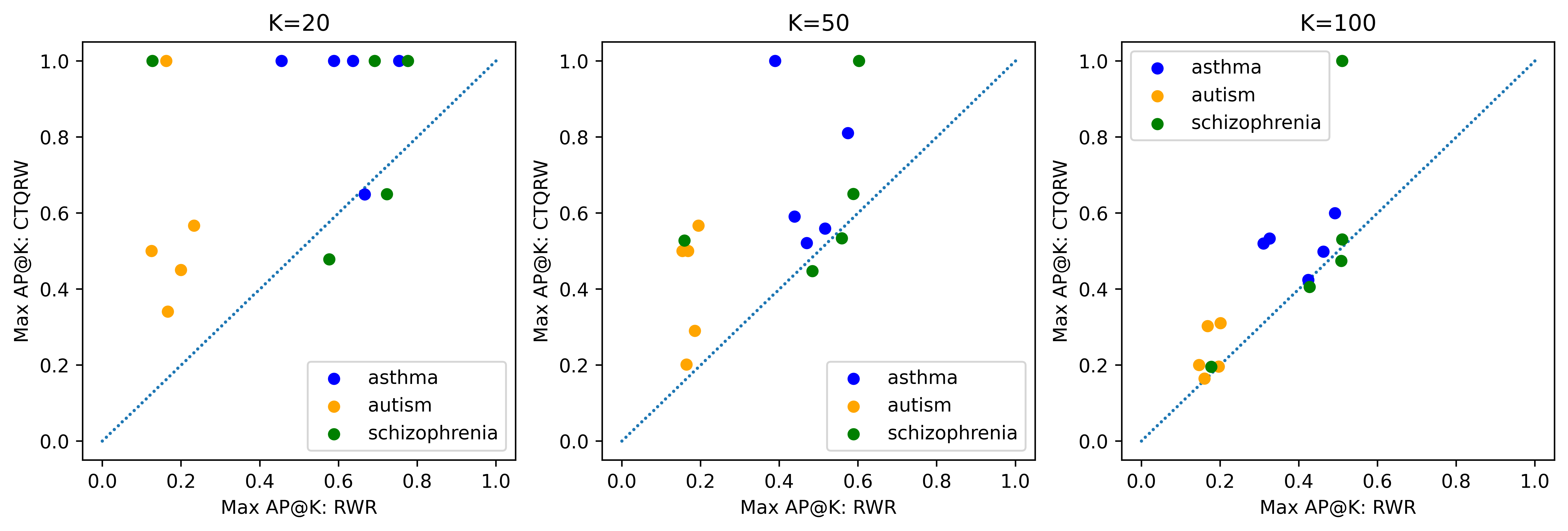}
    \caption{Comparing CTQRW againt RWR across all three diseases and networks with maximum $AP@K$ metric, setting $K=\{20,50,100\}$.}
    \label{fig:ctqrw_v_rwr}
\end{figure*}

We applied CTQRW on Asthma (118 seed genes, 625 target genes), Autism (30 seed genes, 723 target genes), and Schizophrenia (536 seed genes, 949 target genes) as done in Vison\`{a} et al.~\cite{10.1093/bib/bbae014}. We compared the performance of network propagation in the five benchmarking networks of varying sizes (see \cite{10.1093/bib/bbae014} for details). We observed the best performance with respect to mean $AP@K$ in Asthma and Schizophrenia, for both CTQRW and RWR, consistent with previous findings as both diseases had more seed and target genes. The lowest performance was recorded in Autism which also had the least number of seed and target genes (Fig.~\ref{fig:gwas_line}). We observe that CTQRW had far more fluctuations across time, frequently changing mean $AP@K$ with more variance, which also led to it reaching the maximum mean $AP@K$ of 0.7 in HumanNet network for Asthma, 0.3 in PCNet network for Autism, and around 0.55 in BioPlex3 network for Schizophrenia. All of these, except Schizophrenia were significantly better than RWR. Specifically, in the case of Autism the performance is much better in $AP@20$ with CTQRW with a maximum mean of 0.3 whereas the performance with RWR is at 0.14 (Fig.~\ref{fig:gwas_line}). We note that Autism has least number of seed genes to start with and hence the performance of CTQRW is encouraging. When we compared CTQRW with RWR with varying K from 20 to 100, we note that in the maximum metric $AP@K$, CTQRW significantly outperformed RWR for most of the networks in all three diseases (Fig.~\ref{fig:ctqrw_v_rwr}). As we increase $K$ to 50, CTQRW still performs better than RWR, although, the difference between max $AP@50$ decreases. When $K=100$, they perform almost similarly, CTQRW marginally outperforming RWR. As the maximum metric $AP@K$ concerns finding genes with higher ranks in the target nodes, it is important to note that CTQRW manages to find these genes faster and with greater precision than RWR when $K=20$, but as we increase $K$ and find more genes with higher ranks among a larger set, the performance of CTQRW decreases relative to RWR (Fig.~\ref{fig:ctqrw_v_rwr}).


\subsection{Cell-cell Interaction Networks}

\begin{figure}[!htbp]
    \centering
    \includegraphics[width=\linewidth]{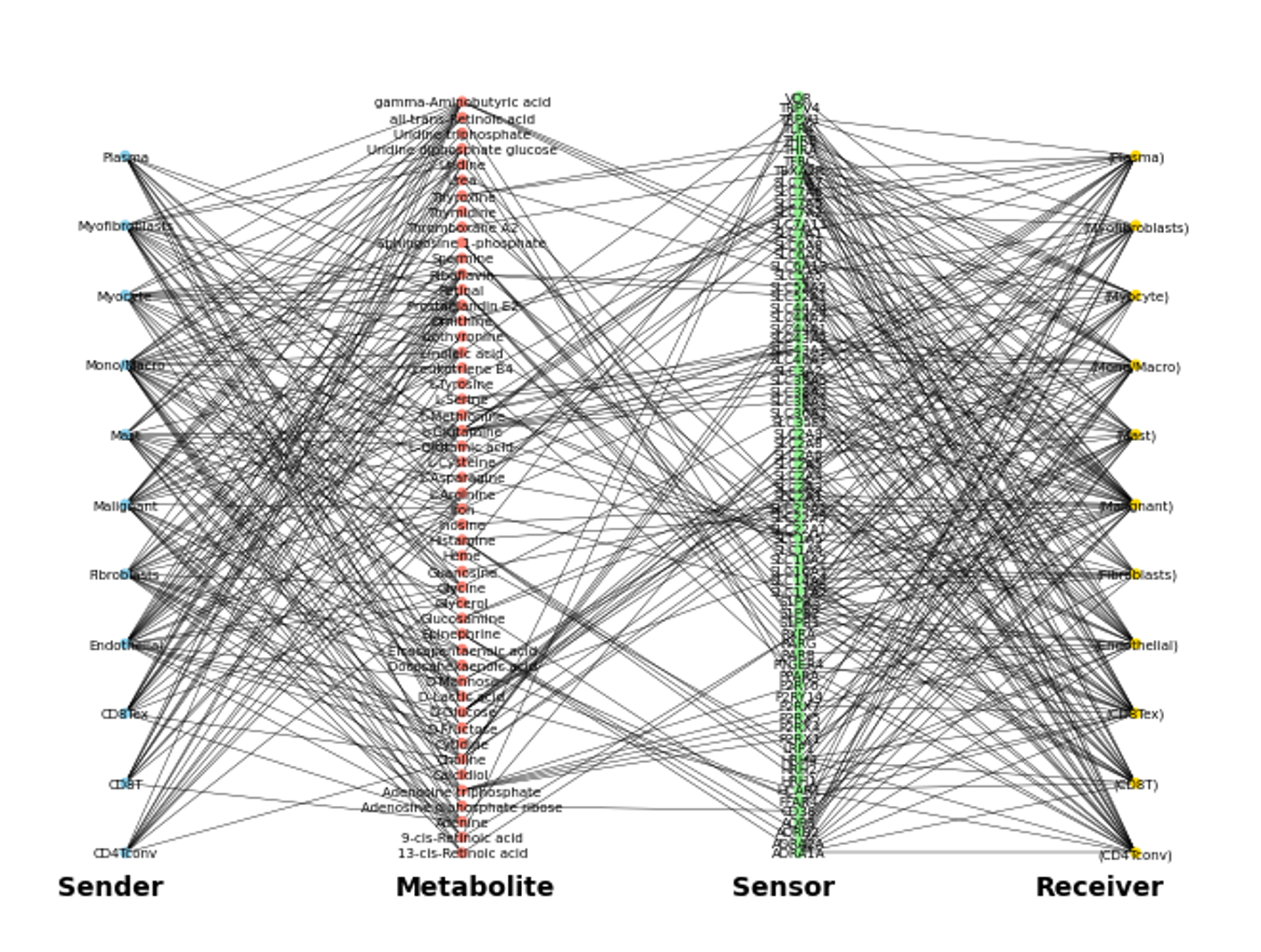}
    \caption{Cell-cell interaction network obtained from single-cell RNA-seq data of mouse brown adipose tissue. It has sender cells and receiver cells on the first and last partition of the graph. The metabolites in the form of ligands span the second partition and the sensor genes in the form of SLC transporters span the third partition. The edges between them are marked in gray.}
    \label{fig:cci_bat}
\end{figure}

\begin{figure}[!b]
    \centering
    \includegraphics[width=\linewidth]{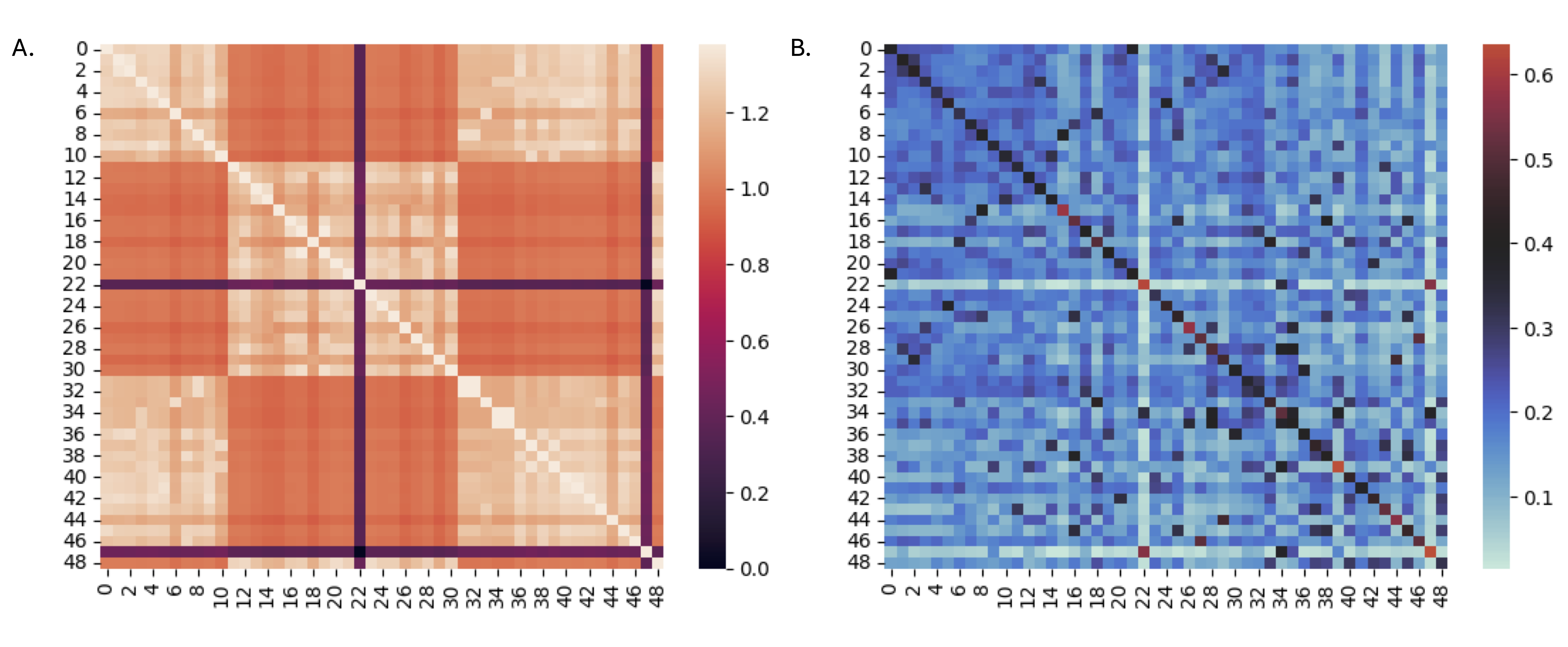}
    \caption{Heatmap of transition probabilities after five steps of the algorithm A. discrete-time random walk and B. discrete-time quantum random walk, respectively. The $l_2$ distance of the transition probability distribution for each pair of vertices $(j,k)$ IS represented in each element of the similarity matrix displayed.}
    \label{fig:heatmap_cci}
\end{figure}

We applied DTQRW to the CCI network (Fig.~\ref{fig:cci_bat}) and compare the results with a classical benchmark in the form of DTRW. Investigating the transition probabilities across all the vertices for all possible edges, we note that the DTRW after five steps of evaluation and corresponding matrix of Euclidean distances between pairs of probability distributions recover five main communities or sub-graphs, between which the walker can jump to.

\begin{figure*}[!htbp]
    \centering
    \includegraphics[width=0.75\linewidth]{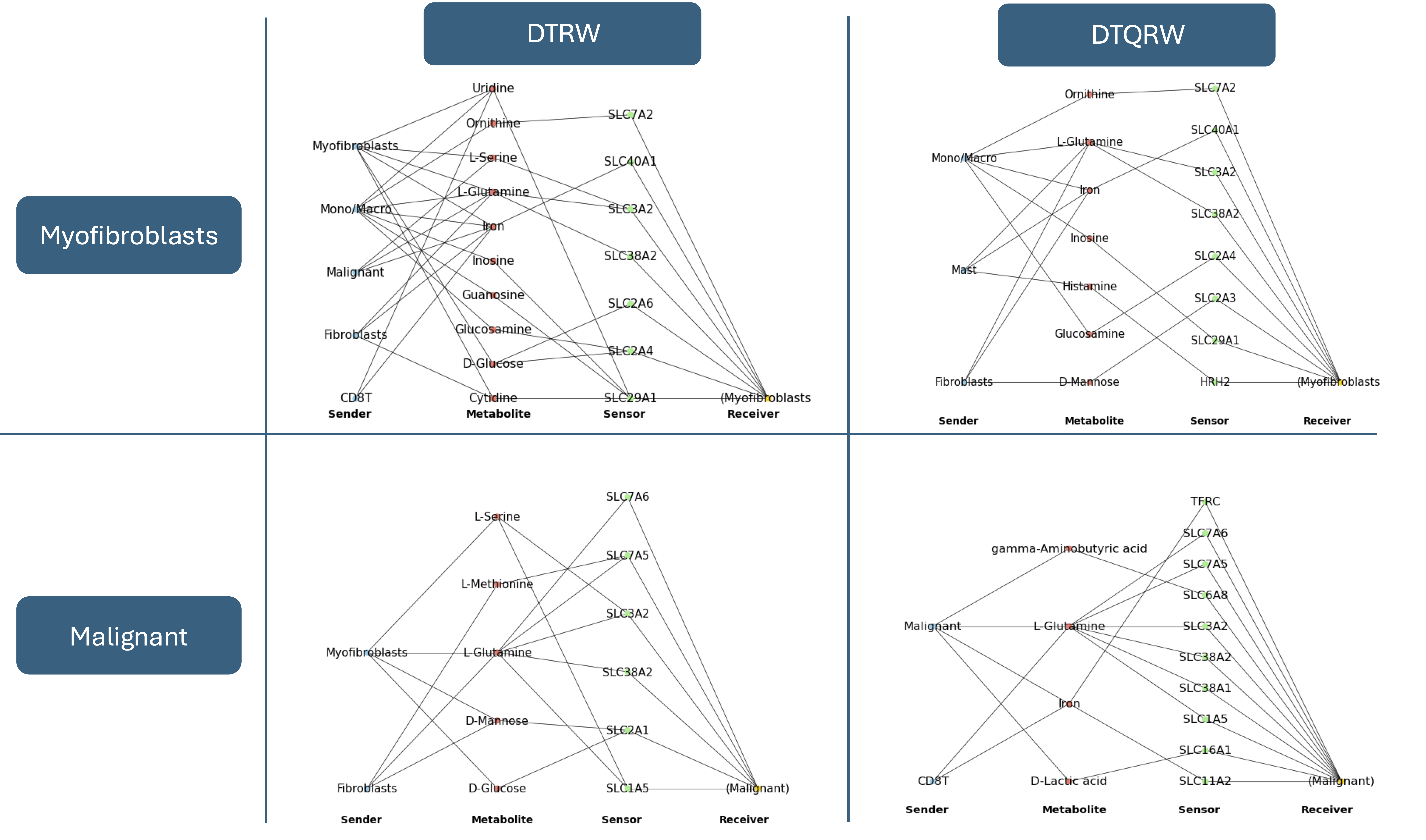}
    \caption{Sub-graph spanned by selecting Myofibroblast and Malignant cells as target nodes and applying DTRW and DTQRW from random source nodes.}
    \label{fig:rw_cci}
\end{figure*}

In the contrary, after applying DTQRW, we observe there were no such well-defined partitions in the graph in transition probabilities and the walker could randomly reach more vertices as the probabilities were randomly high in off-diagonal elements in the similarity matrix, as visible in Fig.~\ref{fig:heatmap_cci}. We note that DTRW induces more nodes and edges than DTQRW when focusing on particular type of cells such as myofibroblasts and malignant cells. We observe that the DTQRW yields a different and shorter community of metabolites than DTRW in both examples (Fig.~\ref{fig:rw_cci}). CD8+ T cells form a key component in the context of cancer progression in adipose tissues~\cite{reina2021cd8+}. 
Obesity triggers metabolic changes in CD8+ T cells that impair their anti-tumor functions. DTRW did not include any path to malignant cells from CD8+T, whereas DTQRW did find a path from CD8+T to malignant cells via L-Glutamine and SLC3A2 sensor (Fig.~\ref{fig:rw_cci}). This is validated in literature as interferon gamma (IFN$\gamma$) released from CD8+T cell downregulates the expression of SLC3A2~\cite{wang2019cd8+} and Glutamine metabolism confers distinct effects on CD8+T cells~\cite{madden2023differential}. 
Gamma-Aminobutyric Acid (GABA) levels are increased in BAT of mice with dietary obesity suggesting that GABA signaling may play a role in adipose tissue metabolism in obesity~\cite{ikegami2018gamma}. DTQRW observes GABA as one of the key metabolites whereas DTRW did not (Fig.~\ref{fig:rw_cci}). In summary, our findings demonstrate that QRW effectively navigate complex biological networks, uncovering informative communities and connections often missed by traditional algorithms. This suggests that DTQRW can potentially unravel how diseases progress within these networks, paving the way for alternative therapeutic discoveries, as quantum random walk methods evolve.


\section{Discussions}
Analysis of complex biological networks remains an open question on which classical and quantum random walks, can provide unique interpretations. Our results show how in the case of three diseases across five biomolecular networks quantum random walks perform better than its classical counterpart on most of the networks and on cell-cell interaction networks, the QRW provided unique insights on communities and relationships within the network that classical random walk might overlook. We also observed many limitations of QRWs emanating from general limitations of the data. Such as, for the particular multipartite structure in the CCI network, we noted that sometimes the DTQRW gets stuck in nodes from which there are no paths to the next partition. The classical random walk also suffers from this issue. However, as reflected in the heatmap of the transition probabilities, the classical random walk is more prone to such dead ends in its path as it sees a specific structure within the network where it can operate (Fig.~\ref{fig:heatmap_cci}). In comparison the DTQRW does not adhere to a defined structure and can transit more freely across the nodes in the multipartite graph. 

We implemented CTQRW in the GWAS seed and target induced biomolecular networks. Our motivation for using continuous- rather than DTQRW is because having a real, rather than an integer-valued hyperparameter $t$, allows us to explore a wider range of probabilities, which is well suited in a random graph, as opposed to a structured graph such as the multi-partite graph in CCI. We observed that some of the networks resulted in fragments,  limiting the paths between seed and target genes, making it difficult to find a target gene in its path when it starts from a seed genes. In particular, autism had more of this challenge as it had a limited number of seed genes to start from. However, we noted that CTQRW was able to find more target genes with higher rank compared to RWR in this setting. Our results indicate that on a variety of biomolecular networks with varying metrics and tasks, QRW can be a promising tool which sees a different Hilbert space when compared to a classical random walk method, due to the inherent principles of quantum mechanics. 

\section{Conclusion}
We explored the application of quantum random walks across diverse biomolecular and cell-cell interaction networks, spanning multiple diseases and single-cell transcriptomic contexts. Our results highlight the remarkable potential of QRW for network propagation tasks, enabling the identification of key genes, network hubs, and biologically relevant pathways. These findings suggest that QRW offer a powerful framework for biomarker discovery and may inform novel therapeutic strategies, opening new avenues for quantum-enhanced biological network analysis.

\section*{Acknowledgment}

The authors thank Ruisheng Wang for discussions in the early stages of this work, which significantly enriched it.

\vspace{12pt}

\bibliographystyle{unsrt}  
\bibliography{refs}
\end{document}